\documentclass[prd,amsfonts,twocolumn,nofootinbib,showpacs]{revtex4}
\RequirePackage{graphicx}
\usepackage{enumerate}

\begin{document}
\title{Modulated decay in the multi-component Universe}

\author{Seishi Enomoto}
\affiliation{Department of Physics, Nagoya University, Nagoya 464-8602,
Japan}
\author{Kazunori Kohri}
\affiliation{Cosmophysics group, Theory Center, IPNS, KEK,
and The Graduate University for Advanced Study (Sokendai),
Tsukuba 305-0801, Japan}
\author{Tomohiro Matsuda}
\affiliation{Laboratory of Physics, Saitama Institute of Technology,
Fukaya, Saitama 369-0293, Japan}

\begin{abstract}
The early Universe after inflation may have oscillations, kinations
 (nonoscillatory 
 evolution of a field), topological defects, 
relativistic and non-relativistic particles at the same time.
The Universe whose energy density is a sum of those components
 can be called the multi-component Universe. 
The components, which may have distinguishable density scalings, may 
decay modulated.
In this paper we study generation of the curvature perturbations 
caused by the modulated decay in the multi-component Universe.
% For our analytic calculation we consider the instant-decay approximation.
% The non-Gaussianity parameter is separated into the
% model-dependent and independent contributions, where the sign of the
% model-independent contribution is determined by the density scalings.
%We also find that the modulated reheating caused by a component whose
% scaling is very close to the radiation may generate significant
% non-Gaussianity even if it is already dominating the Universe.
\end{abstract}

\maketitle

\section{Introduction}
Our focus in this paper is a late-time creation of the curvature
 perturbation $\zeta(k)$, which has cosmological scales beyond the
 horizon when it is created.
The creation  is possible when a mechanism
 works to convert existing isocurvature perturbation of that scale into the 
curvature perturbation. 
To find the creation of the curvature perturbation, we consider  a modulation of decay rate $\Gamma$, which
 is modulated because of the isocurvature perturbation of a moduli. 
Generation of the cosmological perturbations begins presumably during
inflation, when the vacuum fluctuations of light bosonic fields are converted
to a classical perturbation, which gives the seed perturbation (i.e. the
 isocurvature perturbation) that is
 needed for the mechanism~\cite{Lyth-book}. 
Within this general framework, one can find many
proposals~\cite{Lyth-book, Multi1, Multi1a, Multi-NG, modulated-inflation, 
curvaton-paper, hybcurv, infla-curv, PBH-infcurv, IR, end-of-inf, EKM,
IPre, IPhaa}.

First recall the $\delta N$ formalism used to calculate $\zeta$. 
To define the curvature perturbation $\zeta$, the energy density $\rho$
is smoothed on a super-horizon scale shorter than any scale of interest.
One expects this ``separate Universe hypothesis''~\cite{Multi1}
 to be valid for the
calculation, so that one can ensure the maximum regime of applicability
of the calculation.
Then the local energy continuity equation is given by
\begin{equation}
\frac{\partial \rho(x,t) }{\partial t} = - \frac{3}{a(x,t)}
\frac{\partial a(x,t)}
{\partial t} \left( \rho(x,t) + p(x,t) \right)
, \end{equation}
where $t$ is time along a comoving thread of spacetime and $a(t)$ is the local
scale factor. 
During nearly exponential inflation, the vacuum fluctuation of 
each light scalar field $\phi_i$ 
is converted at horizon exit to a nearly Gaussian
classical perturbation with spectrum $(H/2\pi)^2$, where 
the Hubble parameter is $H\equiv \dot a(t)/a(t)$.
 Writing the curvature perturbation
\begin{equation}
\zeta = \delta [ \ln (a(x,t)/a(t_1)] \equiv \delta N
, \end{equation}
 and taking $t_*$ to be an epoch 
during inflation after relevant scales leave the horizon,
we assume $N(\phi_1(x,t_*),\phi_2(x,t_*),\cdots,t,t_*)$ so that 
\begin{equation}
\zeta(x,t) = N_i \delta \phi_i(x,t_*)
+ \frac{1}{2} N_{ij} \delta \phi_i(x,t_*)\delta \phi_j(x,t_*) + \cdots
, \end{equation}
where a subscript $i$ denotes $\partial/\partial \phi_i$ evaluated on the 
unperturbed trajectory.
The $\delta N$-formalism can be applied both during and after inflation.

\begin{figure}[t]
\centering
\includegraphics[width=1.0\columnwidth]{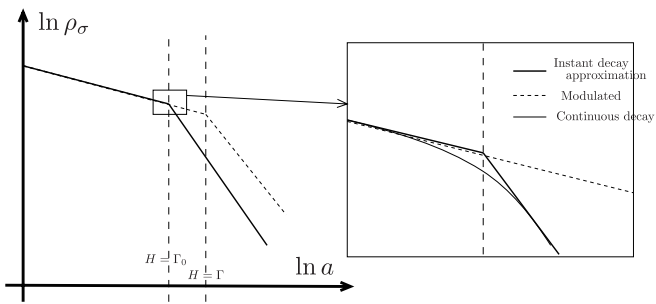}
 \caption{Modulation at the transition causes density perturbations when
the decaying component changes its density scaling.
The straight line shows the instant-decay approximation.} 
\label{fig:instant}
\end{figure}

We consider a density component $\rho_\sigma$ which has a modulated decay rate
$\Gamma(\varphi)$.
Before the decay, $\rho_\sigma$ is not a radiation.
Since we are considering the multi-component Universe, there could be a
radiation background ($\rho_r$) at the same time.
We are not avoiding the case in which $\rho_\sigma$ decays when $\rho_r$
is significant.\footnote{The original scenario of the   
modulated reheating~\cite{IR} assumes that radiation is negligible before
the decay,  
since the ``reheating'' in the original scenario is mostly related to
the inflaton decay.}
Here $\Gamma(\varphi)$ and $\rho_\sigma$ denote the decay rate and the
energy density of the component $\sigma$; and $\Gamma(\varphi)$ is a function
of a moduli $\varphi$ that causes ``modulation''.
Because of the separate Universe hypothesis, the inhomogeneity is
smoothed on a super-horizon scale shorter than any scale of interest.
We also assume instant decay for the calculation~\cite{IR}.
See Fig.\ref{fig:instant} for the basic set-ups of the
modulated reheating scenario and Fig.\ref{fig:mod-uni} for the $\delta
N$ calculation in the separate Universe.

The source of the modulation is the moduli perturbation
of an additional light field $\varphi$, whose potential is
assumed to be negligible at the time of the decay.
The ``seed'' perturbation $\delta \varphi$ is generated during the primordial
inflation.
At the horizon exit, we consider Gaussian perturbation $\delta
\varphi_*\equiv \varphi_*-\bar{\varphi}_*$.
At the decay, we introduce the function
$\varphi=g(\varphi_*)$ and the expansion about the Gaussian
perturbation $\delta \varphi 
\equiv g'\delta \varphi_*$ (or equivalently $\delta \varphi_* =\delta \varphi/g'$).
The function $g$ explains evolution after the horizon
exit~\cite{Lyth-gfc}.\footnote{In our scenario, the evolution of the
moduli does not change the result after the modulated decay.
On the other hand, the late-time evolution of $\varphi$ (moduli) can be
referred to as the famous ``moduli problem''.
We did not consider a specific scenario for the moduli problem,
since the moduli problem is not the target of this paper and the topic
should be separated from the current investigation.}

In our scenario, the first reheating (i.e. the inflaton decay)
occurs before the component
$\rho_\sigma$ decays into radiation.\footnote{Our 
calculation may generically depend on $\delta \rho_\sigma$, which can
cause the curvaton mechanism.
Although we are calculating the modulation when the curvaton mechanism
is negligible, our formalism is carefully prepared so
that the mixed  perturbations can be calculated within the formalism.
See the appendix for more details.}
%%%added after review
The decay of $\rho_\sigma$ may cause secondary ``reheating'' if it is
 dominating the total density at the time of the decay; the secondary
reheating should be discriminated from the first reheating.
In our scenario, ``secondary reheating'' is possible if 
$\rho_\sigma>\rho_r$ at the time when $\rho_\sigma$ decays.
Here $\rho_r$ is the radiation remnant of the first reheating, which may
decrease faster than $\rho_\sigma$.
Therefore,  ``normal modulated reheating'' occurs when
$\rho_\sigma/(\rho_\sigma+\rho_r)=1$, while 
$0.5<\rho_\sigma/(\rho_\sigma+\rho_r)<1$ gives ``near-normal modulated
reheating''.
Finally, $\rho_\sigma/(\rho_\sigma+\rho_r)\le 0.5$ is a modulated decay,
which may not be called ``reheating''.
In any case, ``reheating'' due to $\rho_\sigma$ must be
 distinguished from the conventional reheating.

%%%

For the first example, we consider the simplest two-component Universe,
in which there are $\rho_\sigma\propto a^{-3}$ (matter) and  $\rho_r\propto
a^{-4}$ (radiation) before the decay.
The model is similar to the typical curvaton model, although we are
considering the opposite limit in which the curvaton mechanism is less
significant than the modulation.

Later in this paper we are going to extend our analytic 
calculation to the components that may ``not'' scale like
matter~\cite{IPhaa, NO, Topological-curv};
typical examples are the cosmological defects or the oscillatory 
(could be caused by non-quadratic potentials)/ 
nonoscillatory evolutions~\cite{NO}.

\section{Modulated decay in the simple multi-component Universe}
First we consider the simplest (matter $+$ radiation) multi-component Universe.
In the curvaton mechanism~\cite{curvaton-paper} the significant
contribution comes from the evolution {\bf before} the decay; while the
modulated decay~\cite{IR} describes  
the generation of the curvature perturbations {\bf at the
decay}~\cite{end-of-inf, EKM}.
In the curvaton mechanism, the source of the perturbation is $\delta
\sigma$ ($\delta \rho_\sigma$), while the modulated decay uses
$\delta \varphi$ ($\delta \Gamma$).

For our analytic calculation, we consider instant-decay
approximation~\cite{IR}. 
However, the actual transition could be more complicated depending on
the details 
of the model parameters.
For the modulated reheating scenario, the idea of the continual decay has
been considered by many authors~\cite{Lyth-Malik, prev-inh-dec}.
% In fact, the Universe in which matter is decaying continually into radiation 
% can realize the multi-component Universe during the decay.
% In that way, just by shifting the initial conditions, previous numerical
% studies could have been including ``the modulated 
% decay in the multi-component Universe''.
% In that sense numerical calculation is versatile.
% On the other hand, the analytic estimation, which is calculated in
% specific limits, 
% will help understand the physics behind it. 

% In our analytic calculation we found that model-independent
% non-Gaussianity is significant both in the single and
% in the multi-component Universe.
% The model-dependent non-Gaussianity is calculated from the perturbations
% of $\Gamma (\varphi_*)$.
% Modulated reheating in the curvaton scenario has been considered in
% Ref.\cite{curatonandmodulation}. 
% We hope the reader can easily compare our calculation with those
% previous studies.
\begin{figure}[t]
\centering
\includegraphics[width=1.0\columnwidth]{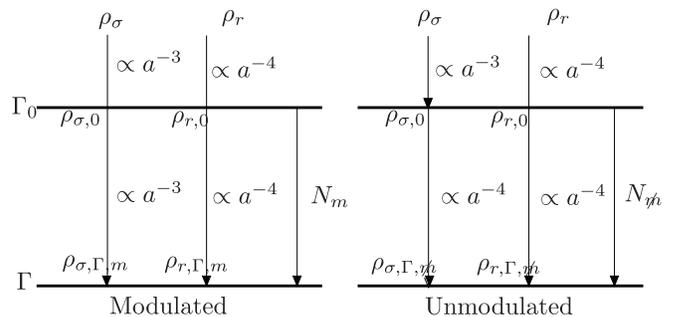}
 \caption{In the left picture we show the densities and their scalings
 in the modulated Universe.
The right picture shows the unmodulated (reference) Universe in which
 the decay occurs at $H=\Gamma_0$.  
{\bf In those pictures we are considering perturbations whose length
 scales are far beyond the horizon size at the time of the decay.
Due to the separate Universe hypothesis,
the inhomogeneity of $\Gamma$ is not explicit in those pictures.}
Note that in the right picture $\rho_{\sigma,\Gamma,\not{m}}$ is the
 radiation created by the  decay of $\rho_\sigma$.
Because of the different $\rho_\sigma$-scalings after $\Gamma_0$, 
one will find $\delta N$ ($\delta N \equiv N_m- N_{\not{m}}$) 
and the difference in the densities at $H=\Gamma$ ($\rho_{i,\Gamma,m}\ne
 \rho_{i,\Gamma,\not{m}}$).}   
\label{fig:mod-uni}
\end{figure}

In our model, the uniform density hypersurface that is 
 defined at the decay is given by
\begin{eqnarray}
\rho_{\sigma,\Gamma} +\rho_{r,\Gamma} &\equiv& 3M_p^2 \Gamma^2,
\end{eqnarray}
where the instant decay occurs at $H=\Gamma$.
More specifically, in the modulated Universe (see the left picture in
Fig.\ref{fig:mod-uni}) we have
\begin{eqnarray}
\rho_{\sigma,\Gamma,m} +\rho_{r,\Gamma,m} &\equiv& 3M_p^2 \Gamma^2,
\end{eqnarray}
and in the reference Universe (see the right picture in
Fig.\ref{fig:mod-uni}) the decay occurs at $H=\Gamma_0$ and we have
at $H=\Gamma$;
\begin{eqnarray}
\rho_{\sigma,\Gamma,\not{m}} +\rho_{r,\Gamma,\not{m}} &\equiv& 3M_p^2 \Gamma^2,
\end{eqnarray}
where $\rho_{\sigma,\Gamma,\not{m}}$ denotes the radiation created by
$\rho_\sigma$. 
In both (modulated and unmodulated) Universe, we define the uniform
density hypersurface at $H=\Gamma_0$ as
\begin{eqnarray}
\rho_{\sigma,0} +\rho_{r,0} &\equiv& 3M_p^2 \Gamma^2_0.
\end{eqnarray}
Without loss of generality, one may choose $\Gamma <\Gamma_0$ for the
calculation.

Using the density scalings, we find in the modulated Universe
\begin{eqnarray}
\rho_{\sigma,\Gamma,m}&=&\rho_{\sigma,0}\left(\frac{a_{\Gamma,m}}{a_{\Gamma_0}}\right)^{-3}
\nonumber\\
\rho_{r,\Gamma,m}&=&\rho_{r,0}\left(\frac{a_{\Gamma,m}}{a_{\Gamma_0}}\right)^{-4},
\end{eqnarray}
which lead to
\begin{equation}
\label{above-01}
\frac{\rho_{\sigma,0}\left(\frac{a_{\Gamma,m}}{a_{\Gamma_0}}\right)^{-3}
 +\rho_{r,0}\left(\frac{a_{\Gamma,m}}{a_{\Gamma_0}}\right)^{-4}}
{\rho_{\sigma,0} +\rho_{r,0}}=\frac{\Gamma^2}{\Gamma_0^2}.
\end{equation}
Defining ``$N_m$'' in the modulated Universe as 
\begin{equation}
N_m\equiv \int_{t_{\Gamma_0}}^{t_{\Gamma}}H(t)dt,
\end{equation}
where the subscripts $\Gamma$ and $\Gamma_0$ denote the hypersurfaces
$H=\Gamma$ and $H=\Gamma_0$, one can rewrite Eq.(\ref{above-01}) as
\begin{eqnarray}
\label{mod-N}
f_\sigma e^{-3N_m}+ (1-f_\sigma) e^{-4N_m}
&=&\frac{\Gamma^2}{\Gamma_0^2},
\end{eqnarray}
where the coefficient is defined by
\begin{eqnarray}
f_\sigma &\equiv& \frac{\rho_{\sigma,0}}
{\rho_{\sigma,0} +\rho_{r,0}}.
\end{eqnarray}

In order to compare $N_m$ with the unmodulated Universe, 
we find a similar equation in the unmodulated Universe,
\begin{eqnarray}
\label{unmod-N}
f_\sigma e^{-4N_{\not{m}}}+ (1-f_\sigma) e^{-4N_{\not{m}}}
&=&\frac{\Gamma^2}{\Gamma_0^2}.
\end{eqnarray}

If one needs to understand the relation between the modulated decay and
 the curvaton mechanism, the curvaton density perturbation 
$\delta \rho_\sigma$ must be included at $H=\Gamma_0$. 
Here we consider non-linear formalism of
 Ref.\cite{Lyth-general,Langlois:2008vk}.\footnote{A more straight
 definition of the curvaton/modulation in the light of 
 the $\delta N$ formalism can be found in Ref.\cite{Chiamin-mod}.}
We find that the component perturbations are defined as
\begin{eqnarray}
\label{eq-14}
\zeta_{\sigma}&=&\delta
 N_\mathrm{ini}+\frac{1}{3}\int^{\rho_{\sigma,0}}_{\bar{\rho}_{\sigma,0}} 
\frac{d\tilde{\rho}_\sigma}{\tilde{\rho}_\sigma}\\
%\zeta_{\sigma,\not{m}}&=&\delta
%N_\mathrm{ini}+\frac{1}{4}\int^{\rho_{\sigma,0}}_{\bar{\rho}_{\sigma,0}} 
%\frac{d\tilde{\rho}_\sigma}{\tilde{\rho}_\sigma}
\label{eq-15}
\zeta_r&=&\delta N_\mathrm{ini}+\frac{1}{4}
\int^{\rho_{r,0}}_{\bar{\rho}_{r,0}}
\frac{d\tilde{\rho}_r}{\tilde{\rho}_r},
\end{eqnarray}
where $\rho_{\sigma,0}$ and $\rho_{r,0}$ are defined at $H=\Gamma_0$, 
and $\bar{\rho}_i$ denotes their mean value.
$\tilde{\rho}_i$ does not define a new quantity, but is just
introduced to define the integral.
Here $\delta N_\mathrm{ini}$ denotes the curvature perturbation
before the curvaton mechanism, which is usually neglected in the
conventional curvaton calculation.
Finally, the non-linear formalism gives 
\begin{eqnarray}
\rho_{\sigma,0}&=&\bar{\rho}_{\sigma,0}e^{3(\zeta_{\sigma}-\delta
 N_\mathrm{ini})}\\
%\rho_{\sigma,0}&=&\bar{\rho}_{\sigma,0}e^{4(\zeta_{\sigma,\not{m}}-\delta
% N_\mathrm{ini})}\nonumber\\
\rho_{r,0}&=&\bar{\rho}_{\sigma,0}e^{4(\zeta_{r}-\delta
 N_\mathrm{ini})},
\end{eqnarray}
which leads to 
\begin{eqnarray}
f_{\sigma} &=& \frac{\bar{\rho}_{\sigma,0}e^{3(\zeta_{\sigma}-\delta
 N_\mathrm{ini})}}
{\bar{\rho}_{\sigma,0}e^{3(\zeta_{\sigma}-\delta N_\mathrm{ini})}
 +\bar{\rho}_{r,0}e^{4(\zeta_{r}-\delta N_\mathrm{ini})}}\nonumber\\
&=& \frac{\bar{\rho}_{\sigma,0}e^{3(\zeta_{\sigma}-\delta
 N_\mathrm{ini})}}
{3M_p^2 \Gamma_0^2}.
%f_{\sigma,\not{m}} &\equiv& \frac{\bar{\rho}_{\sigma,0}
%e^{4(\zeta_{\sigma,\not{m}}-\delta
% N_\mathrm{ini})}}
%{\rho_{\sigma,0}e^{4(\zeta_{\sigma,\not{m}}-\delta N_\mathrm{ini})}
% +\rho_{r,0}e^{4(\zeta_{r}-\delta N_\mathrm{ini})}}\nonumber\\
%&=&\frac{\bar{\rho}_{\sigma,0}
%e^{4(\zeta_{\sigma,\not{m}}-\delta
% N_\mathrm{ini})}}
%{3M_p^2 \Gamma_0^2}.
\end{eqnarray}
Using the above equation, we
find from Eq.(\ref{mod-N}) and (\ref{unmod-N});
\begin{eqnarray}
\label{motonoshiki}
\bar{f}_{\sigma} e^{3(\zeta_{\sigma}-\delta N_\mathrm{ini}-N_m)}+ (1-\bar{f}_{\sigma})
 e^{4(\zeta_r-\delta N_\mathrm{ini}-N_m)}
&=&\frac{\Gamma^2}{\Gamma_0^2}\nonumber\\
\bar{f}_{\sigma} e^{3(\zeta_{\sigma}-\delta N_\mathrm{ini})-4N_{\not{m}}}
+ (1-\bar{f}_{\sigma}) e^{4(\zeta_r-\delta N_\mathrm{ini}-N_{\not{m}})}
&=&\frac{\Gamma^2}{\Gamma_0^2},\nonumber\\
\end{eqnarray}
where the coefficient is defined by
\begin{eqnarray}
\bar{f}_\sigma &\equiv& \frac{\bar{\rho}_{\sigma,0}}
{3M_p^2 \Gamma_0^2}.
\end{eqnarray}

\subsection{First order}
If a function $G$ is perturbed, one can expand
\begin{equation}
G=\bar{G}+\sum_{k=1}^{\infty}\frac{1}{k!}\delta G^{(k)}.
\end{equation}
Therefore, from Eq.(\ref{motonoshiki}), we find at first order 
\begin{eqnarray}
2\frac{\delta
 \Gamma^{(1)}_m}{\Gamma_0}&=&3\bar{f}_\sigma(\zeta_{\sigma}^{(1)}-\delta
 N_\mathrm{ini}^{(1)}- 
 N_m^{(1)})\nonumber\\
&&+4(1-\bar{f}_\sigma)(\zeta_{r}^{(1)}-\delta N_\mathrm{ini}^{(1)}
-N_m^{(1)})\\
&& -9\bar{f}_\sigma N_m^{(0)}(\zeta_{\sigma}^{(1)}-\delta
 N_\mathrm{ini}^{(1)})\nonumber\\
&&-16(1-\bar{f}_\sigma)N_m^{(0)}(\zeta_{r}^{(1)}-\delta
 N_\mathrm{ini}^{(1)})\\
2\frac{\delta\Gamma^{(1)}_{\not{m}}}{\Gamma_0}
&=&\bar{f}_\sigma(3\zeta_{\sigma}^{(1)}-3\delta 
N_\mathrm{ini}^{(1)}-4N_{\not{m}}^{(1)})
\nonumber\\
&&+4(1-\bar{f}_\sigma)(\zeta_{r}^{(1)}-\delta N_\mathrm{ini}^{(1)}
- N_{\not{m}}^{(1)})\nonumber\\
&& -12\bar{f}_\sigma N_{\not{m}}^{(0)}(\zeta_{\sigma}^{(1)}-\delta
 N_\mathrm{ini}^{(1)})\nonumber\\
&&-16(1-\bar{f}_\sigma)N_{\not{m}}^{(0)}(\zeta_{r}^{(1)}-\delta
 N_\mathrm{ini}^{(1)}),
\end{eqnarray}
where the curvaton mechanism\footnote{See also the appendix to
 understand the definition of ``curvaton mechanism'' used above.}
 between $H=\Gamma_0$ and $H=\Gamma$
 vanishes at this 
 order, because we have a trivial relation $N_m^{(0)}=N_{\not{m}}^{(0)}=0$.

Solving the above equations, we find
\begin{eqnarray}
N^{(1)}_m&=& -p_\sigma \frac{\delta \Gamma^{(1)}_m}{\Gamma_0}
+r_\sigma (\zeta_{\sigma}^{(1)}-\delta N_\mathrm{ini}^{(1)})\nonumber\\
&&+(1-r_\sigma)(\zeta_r^{(1)}-\delta N_\mathrm{ini}^{(1)})\\
%&&-3N_m^{(0)}r_\sigma(\zeta_{\sigma}^{(1)}-\delta
% N_\mathrm{ini}^{(1)})\nonumber\\
%&&-4N_m^{(0)}r_\sigma(\zeta_{r}^{(1)}-\delta
% N_\mathrm{ini}^{(1)})\\
N_{\not{m}}^{(1)}&=& -\frac{1}{2} \frac{\delta
 \Gamma^{(1)}_{\not{m}}}{\Gamma_0}+\frac{3}{4}\bar{f}_\sigma (\zeta_{\sigma}^{(1)}-\delta
 N_\mathrm{ini}^{(1)})\nonumber\\
&&+(1-\bar{f}_\sigma)(\zeta_r^{(1)}-\delta
 N_\mathrm{ini}^{(1)}),
%&&-4N_{\not{m}}^{(0)}r_\sigma(\zeta_{\sigma}^{(1)}-\delta
% N_\mathrm{ini}^{(1)})\nonumber\\
%&&-4N_{\not{m}}^{(0)}r_\sigma(\zeta_{r}^{(1)}-\delta
% N_\mathrm{ini}^{(1)}),
\end{eqnarray}
where the coefficients are defined by
\begin{eqnarray}
\label{eq-27}
p_\sigma&\equiv& \frac{2(\bar{\rho}_{\sigma,0}+\bar{\rho}_{r,0})}
{3\bar{\rho}_{\sigma,0}+4\bar{\rho}_{r,0}}\nonumber\\
r_\sigma &\equiv& \frac{3 \bar{\rho}_{\sigma,0}}
{3 \bar{\rho}_{\sigma,0}+4 \bar{\rho}_{r,0}}.
\end{eqnarray}
Using the above equations, we find the relation
\begin{equation}
p_\sigma-\frac{1}{2}=\frac{1}{6}r_\sigma.
\end{equation}
We have to calculate $N_{\not{m}}$ since $\delta N$ measures 
the deviation from the reference Universe.
Therefore, the curvature perturbation created by the modulation
($\delta N^{(1)}\equiv N_m^{(1)}-N_{\not{m}}^{(1)}$) is
calculated as 
\begin{eqnarray}
\delta N^{(1)}&=& \left(-p_\sigma\frac{\delta\Gamma^{(1)}_m}{\Gamma_0}
+\frac{1}{2}\frac{\delta\Gamma^{(1)}_{\not{m}}}{\Gamma_0}\right),
\end{eqnarray}
where other terms cancel by definition.\footnote{
From Eq.(\ref{eq-27}), we find
\begin{eqnarray}
r_\sigma-\frac{3}{4}f_\sigma&=&\frac{3 \bar{\rho}_{\sigma,0}}
{3 \bar{\rho}_{\sigma,0}+4 \bar{\rho}_{r,0}}
-\frac{3}{4}
\frac{\bar{\rho}_{\sigma,0}}
{\bar{\rho}_{\sigma,0}+\bar{\rho}_{r,0}}\nonumber\\
&=&\frac{3}{4}\bar{\rho}_{\sigma,0}\times\left[
\frac{\bar{\rho}_{\sigma,0}}
{(3\bar{\rho}_{\sigma,0}+4\bar{\rho}_{r,0})
(\bar{\rho}_{\sigma,0}+\bar{\rho}_{r,0})}\right],\\
f_\sigma-r_\sigma&=&\frac{\bar{\rho}_{\sigma,0}}
{\bar{\rho}_{\sigma,0}+\bar{\rho}_{r,0}}
-\frac{3 \bar{\rho}_{\sigma,0}}
{3 \bar{\rho}_{\sigma,0}+4 \bar{\rho}_{r,0}}
\nonumber\\
&=&\bar{\rho}_{r,0}\times\left[
\frac{\bar{\rho}_{\sigma,0}}
{(3\bar{\rho}_{\sigma,0}+4\bar{\rho}_{r,0})
(\bar{\rho}_{\sigma,0}+\bar{\rho}_{r,0})}\right].
\end{eqnarray}
Therefore we have the relation
\begin{eqnarray}
r_\sigma-\frac{3}{4}f_\sigma&=&(f_\sigma-r_\sigma)\times \frac{3}{4}
\frac{\bar{\rho}_{\sigma,0}}{\bar{\rho}_{r,0}}
\end{eqnarray}
From Eq.(\ref{eq-14}) and (\ref{eq-15}), we find
\begin{eqnarray}
\zeta_{\sigma}-\delta
 N_\mathrm{ini}&=&\frac{1}{3}\int^{\rho_{\sigma,0}}_{\bar{\rho}_{\sigma,0}} 
\frac{d\tilde{\rho}_\sigma}{\tilde{\rho}_\sigma}\\
\zeta_r-\delta N_\mathrm{ini}&=&\frac{1}{4}
\int^{\rho_{r,0}}_{\bar{\rho}_{r,0}}
\frac{d\tilde{\rho}_r}{\tilde{\rho}_r},
\end{eqnarray}
which give (at first order)
\begin{eqnarray}
\zeta_{\sigma}^{(1)}-\delta N_\mathrm{ini}^{(1)}
&=&\frac{1}{3}
\frac{\delta \rho_\sigma}{\bar{\rho}_\sigma}\\
\zeta_r^{(1)}-\delta N_\mathrm{ini}^{(1)}&=&\frac{1}{4}
\frac{\delta \rho_r}{\bar{\rho}_r}.
\end{eqnarray}
Therefore, we find that the terms $\left(r_\sigma-\frac{3}{4}f_\sigma\right)
\left(\zeta_\sigma-N_\mathrm{ini}\right)$ and
$\left(f_\sigma-r_\sigma\right)\left(\zeta_r-N_\mathrm{ini}\right)$
cancels because of the relation
\begin{eqnarray}
\left(r_\sigma-\frac{3}{4}f_\sigma\right)
\left(\zeta_\sigma-N_\mathrm{ini}\right) 
&=&\left[(f_\sigma-r_\sigma)\times \frac{3}{4}
\frac{\bar{\rho}_{\sigma,0}}{\bar{\rho}_{r,0}}\right]
\times \left[\frac{1}{3}
\frac{\delta \rho_\sigma}{\bar{\rho}_\sigma}\right]\nonumber\\
&=&\left[(f_\sigma-r_\sigma)\right]
\times \left[\frac{1}{4}
\frac{\delta \rho_\sigma}{\bar{\rho}_r}\right]\nonumber\\
&=&
-\left(f_\sigma-r_\sigma\right)\left(\zeta_r-N_\mathrm{ini}\right).
\end{eqnarray}
Here the last line is obtained using $\delta \rho_\sigma +\delta
      \rho_r=\delta \rho_\mathrm{tot}\equiv 0$.}

Now we consider the perturbation of
$\Gamma$ with respect to the modulation.
Expanding $\varphi_{d}\equiv g(\varphi_*)$, which defines
$\varphi$ at the decay,
we find~\cite{SVW}
\begin{eqnarray}
\varphi_d&=&\bar{g}+\delta \varphi.
\end{eqnarray}
where $\delta \varphi\equiv g'\delta \varphi_*$.
In that way the first order perturbation of the decay rate is
calculated as 
\begin{equation}
\delta \Gamma^{(1)}=\left[\frac{\partial \Gamma}{\partial
		     g}\right]_{g=\bar{g}}
\delta \varphi.
\end{equation}
In the practical calculation $\delta \Gamma_m$ and $ \delta
\Gamma_{\not{m}}$ are identical.
We thus find 
\begin{eqnarray}
\delta N^{(1)}&=& \left(-p_\sigma+\frac{1}{2}\right)
\frac{\delta\Gamma^{(1)}}{\Gamma_0},
\end{eqnarray}
where $\delta \Gamma^{(1)}\equiv \delta \Gamma^{(1)}_m=\delta \Gamma^{(1)}_{\not{m}}$.
In the single-component limit ($p_\sigma=2/3$), we find
\begin{eqnarray}
\delta N^{(1)}&=& -\frac{1}{6}
\frac{\delta\Gamma^{(1)}}{\Gamma_0}\nonumber\\
&=& -\frac{1}{6}\frac{\Gamma'}{\Gamma_0}\delta \varphi,
\end{eqnarray}
which reproduces the calculation in Ref.\cite{IR}.

It is obvious that $g$ is trivial in the slow-roll limit; however
for more practical estimation one might have to calculate the function $g$,
which can depend on the details of the model and the cosmological evolutions.

\subsection{Second order}
Generically, one can expand
\begin{equation}
\varphi=\bar{\varphi}+\sum_{k=1}^{\infty}\frac{1}{k!}\delta\varphi^{(k)},
\end{equation}
where $\delta\varphi^{(1)}$ is a Gaussian random field.
In the same way, the primordial perturbation can be expanded as
\begin{equation}
\zeta=\zeta^{(1)}+\sum_{k=2}^{\infty}\frac{1}{k!}
\zeta^\mathrm{(k)},
\end{equation}
where $\zeta^{(1)}$ is Gaussian.
Non-linearity parameters are defined for the adiabatic perturbation $\zeta$;
\begin{equation}
\zeta=\zeta^{(1)}+\frac{3}{5}f_{NL}(\zeta^{(1)})^2+\frac{9}{25}g_{NL}
(\zeta^{(1)})^3+.... 
\end{equation}

Using the Gaussian quantum fluctuations at the horizon exit ($\delta
\varphi_{*}$), we can write~\cite{SVW} 
\begin{equation}
\varphi_{*}=\bar{\varphi}_{*}+\delta\varphi_{*},
\end{equation}
which is exact by definition.
Again, we write 
\begin{equation}
\varphi_{d}\equiv g(\varphi_{*})
\end{equation}
and expand it as~\cite{SVW}
\begin{eqnarray}
\varphi_\mathrm{ini}&=&\bar{g}+\sum_{k=1}^{\infty}
 \frac{1}{k!}g^{(k)}\left(\frac{\bar{g}}{g'}\frac{\delta\varphi}
{\bar{\varphi}}
\right)^k,
\end{eqnarray}
where we wrote $g^{(k)}\equiv \partial^k g/\partial \varphi_{*}^k$.

\subsubsection{Decay rates}
Before discussing non-Gaussianity of the second order perturbations,
we consider the expansion of the decay rate for some specific examples.
\begin{itemize}
\item Our first example is
\begin{equation}
\Gamma (\varphi)=\bar{\Gamma}
\left(1 +\frac{1}{2}\frac{\varphi^2}{M_*^2}\right).
\end{equation}
Then, one can expand
\begin{equation}
\frac{\Gamma (\varphi_d)}{\bar{\Gamma}}=1  +\frac{1}{2}\frac{
\left[
\bar{g}+\sum_{k=1}^{\infty}
 \frac{1}{k!}g^{(k)}\left(\frac{\bar{g}}{g'}\frac{\delta\varphi}{\bar{\varphi}}
\right)^k
\right]^2
}{M_*^2}.
\end{equation}
We thus find for the expansion $\Gamma=\Gamma_0+\delta
\Gamma^{(1)}+\frac{1}{2}\Gamma^{(2)}+...$ with the approximation
$\bar{\Gamma}\simeq \Gamma_0$;
\begin{eqnarray}
\frac{\delta \Gamma^{(1)}}{\Gamma_0}&\equiv&
\frac{\bar{g}}{M_*^2}\delta \varphi\\
\frac{\delta \Gamma^{(2)}}{\Gamma_0}&\equiv&
\frac{1}{M_*^2}\left[1+\frac{g''\bar{g}}{(g')^2}\right](\delta
\varphi)^2\nonumber\\
&=&\frac{M_*^2}{\bar{g}^2}\left[1+\frac{g''\bar{g}}{(g')^2}\right]
\left(\frac{\delta \Gamma^{(1)}}{\Gamma_0}\right)^2.
\end{eqnarray}

An interesting case would be $g\lesssim M_*$, where the initial
      condition is comparable but less than the cut-off scale.
In that case one can find significant $f_{NL}$ in the conceivable range.
Moreover, it is possible to find negative contribution from
\begin{equation}
\Gamma (\varphi)=\bar{\Gamma}
\left(1 -\frac{1}{2}\frac{\varphi^2}{M_*^2}\right).
\end{equation}
The flip of the sign is very important.

\item Second, we consider $\Gamma\propto \varphi^n$.
The specific form becomes
\begin{equation}
\Gamma (\varphi)=\frac{\lambda^{(n)}}{n}\frac{\varphi^n}{M_*^{n-1}}.
\end{equation}
Then, $\Gamma$ can be expanded as 
\begin{eqnarray}
\frac{\delta \Gamma^{(1)}}{\Gamma_0}&=& 
\left[n \frac{\delta \varphi}{\bar{g}}\right]\\
\frac{\delta \Gamma^{(2)}}{\Gamma_0}&=&
\frac{1}{2}\left[(n-1)+\frac{\bar{g} g''}{(g')^2}\right] 
\left(\frac{\delta \Gamma^{(1)}}{\Gamma_0}\right)^2.
\end{eqnarray}
\end{itemize}

Let us summarize the results.
Defining 
\begin{eqnarray}
\label{gammma2def}
\frac{\delta \Gamma^{(2)}}{\Gamma_0}&=&
A \left(\frac{\delta \Gamma^{(1)}}{\Gamma_0}\right)^2,
\end{eqnarray}
we find 
\begin{eqnarray}
A&=&\pm \frac{M_*^2}{\bar{g}^2}\left[1+\frac{g''\bar{g}}{(g')^2}\right]\nonumber\\
&\mathrm{for}&\Gamma (\varphi)=\Gamma_* \left(1
			   \pm\frac{1}{2}\frac{\varphi^2}{M_*^2}\right),
\end{eqnarray}
and
\begin{eqnarray}
A&=&\frac{1}{2}\left[(n-1)+\frac{\bar{g} g''}{(g')^2}\right]\nonumber\\
&\mathrm{for}&
\Gamma
 (\varphi)=\frac{\lambda^{(n)}}{n}\frac{\varphi^n}{M_*^{n-1}},
\end{eqnarray}
where ``$A$'' is determined by $\Gamma(\varphi)$ and $g$.

The above results are considered when we estimate 
the non-Gaussianity parameter $f_{NL}$.

\subsubsection{$f_{NL}$}
In order to extract the contributions from the modulation,
we are going to assume $\zeta_\sigma\simeq \zeta_r\simeq 0$.
We also assume $\delta N_\mathrm{ini}\simeq 0$ for simplicity.

Then, one can easily expand Eq.(\ref{motonoshiki}) to find the second
order perturbations.
The expansions used here are
\begin{eqnarray}
e^{aN}&=& 1+ a(N^{(1)}+\frac{1}{2}N^{(2)}+...)\nonumber\\
&&+\frac{a^2}{2}(N^{(1)}+\frac{1}{2}N^{(2)}+...)^2+...,
\end{eqnarray}
and
\begin{equation}
\left(\frac{\Gamma}{\Gamma_0}\right)^2
=\left(\frac{\Gamma_0+\delta \Gamma^{(1)}+\frac{1}{2}\delta
  \Gamma^{(2)}+...}{\Gamma_0}\right)^2.
\end{equation}
We find for the second order perturbations
\begin{eqnarray}
N_m^{(2)}&=& p_\sigma
\left[\left(8-\frac{7}{2}\bar{f}_\sigma\right)\left(N_m^{(1)}\right)^2
-\frac{\left(\delta \Gamma^{(1)}\right)^2+\Gamma_0\delta \Gamma^{(2)}}
{\Gamma_0^2}\right]\nonumber\\
N_{\not{m}}^{(2)}&=& \frac{1}{2}
\left[8\left(N_{\not{m}}^{(1)}\right)^2
-\frac{\left(\delta \Gamma^{(1)}\right)^2+\Gamma_0\delta \Gamma^{(2)}}
{\Gamma_0^2}\right].
\end{eqnarray}
Using the relations between the first order perturbations;
\begin{eqnarray}
N_m^{(1)}&=&-p_\sigma \frac{\delta \Gamma^{(1)}}{\Gamma_0}\nonumber\\
&=&- \frac{2p_\sigma}{1-2p_\sigma}\delta N^{(1)}\\
N_{\not{m}}^{(1)}&=&-\frac{1}{2}\frac{\delta \Gamma^{(1)}}{\Gamma_0}\nonumber\\
&=&- \frac{1}{1-2p_\sigma}\delta N^{(1)},
\end{eqnarray}
and the definition (\ref{gammma2def}), we find 
\begin{eqnarray}
N_m^{(2)}&=& \left[
\frac{p_\sigma^3(32-14\bar{f}_\sigma)}{(1-2p_\sigma)^2}
-\frac{4p_\sigma(1+A)}{(1-2p_\sigma)^2}\right]\left(\delta
N^{(1)}\right)^2
\nonumber\\
N_{\not{m}}^{(2)}&=& \left[
\frac{4}{(1-2p_\sigma)^2}
-\frac{2(1+A)}{(1-2p_\sigma)^2}\right]\left(\delta
N^{(1)}\right)^2.\nonumber\\
\end{eqnarray}
We thus find 
\begin{eqnarray}
f_{NL}&=&\frac{5}{3}
\left[\frac{p_\sigma^3(16 -7\bar{f}_\sigma)-2p_\sigma-1}{(1-2p_\sigma)^2}
\right]\nonumber\\
&&+\frac{5A}{3}\frac{1}{1-2p_\sigma},
\end{eqnarray}
where the last term depends on $A$.
In the single-component limit ($\bar{f}_\sigma \rightarrow 1$ and
$p_\sigma\rightarrow 2/3$), we find a simple formula
\begin{eqnarray}
f_{NL}&=&5 -5A.
\end{eqnarray}
Note that the A-independent contribution  $f_{NL}=5$ in the ``normal reheating
limit'' is showing an interesting result.
Note also that $\Gamma\propto \varphi^3$ gives $A\sim 1$ when $g$ 
is trivial and it lead to the cancellation ($f_{NL}\simeq 0$).

In the opposite limit, $\bar{f}_\sigma\rightarrow 0$ leads to
$p_\sigma\rightarrow 1/2$. 
In that limit we find
\begin{equation}
f_{NL}\propto \frac{1}{(1-2p_\sigma)^2}\gg 1.
\end{equation}

\section{Higher potential or topological defects}
For the scalar potential of the form $V(\sigma)\propto\sigma^n$,
the energy density of the scalar-field oscillations decreases as
$\rho_\sigma\propto a^{\frac{-6n}{n+2}}$ when the oscillations are rapid
compared with the expansion rate~\cite{coherent-turner}.
Alternatively, one may choose topological defects for the decaying
component, which may scale like $\rho_\sigma \propto a^{k}$. 
NO (Non Oscillatory) motion can lead to a different density scaling~\cite{NO}.
Here the scaling is approximately 
defined at the time of the decay;
there is no need to find exact scale-dependence that is valid during the
whole evolution.
This point might be crucial for the practical investigation.

For our purpose, we consider the component that scales like
$\rho_\sigma\propto a^{-(4+\epsilon_n)}$.
Here $\epsilon_n=-1$ corresponds to the sinusoidal oscillation for the 
quadratic potential, whose energy density scales like
$\rho_\sigma\propto a^{-3}$.
Note that $\epsilon_n\ge 0$ is not excluded in our calculation;
we will show that this may change the sign of $f_{NL}$.

In order to include the isocurvature perturbation at $H=\Gamma_0$,
we consider the component perturbations defined by
\begin{eqnarray}
\zeta_{\sigma}&=&\delta
 N_\mathrm{ini}+\frac{1}{4+\epsilon_n}
\int^{\rho_{\sigma,0}}_{\bar{\rho}_{\sigma,0}} 
\frac{d\tilde{\rho}_\sigma}{\tilde{\rho}_\sigma}\\
\zeta_r&=&\delta N_\mathrm{ini}+\frac{1}{4}
\int^{\rho_{r,0}}_{\bar{\rho}_{r,0}}
\frac{d\tilde{\rho}_r}{\tilde{\rho}_r}.
\end{eqnarray}
Then  we find
\begin{eqnarray}
\label{motonoshiki2}
\bar{f}_{\sigma} e^{(4+\epsilon_n)
(\zeta_{\sigma}-\delta N_\mathrm{ini}-N_m)}+ (1-\bar{f}_{\sigma})
 e^{4(\zeta_r-\delta N_\mathrm{ini}-N_m)}
&=&\frac{\Gamma^2}{\Gamma_0^2}\nonumber\\
\bar{f}_{\sigma} e^{(4+\epsilon_n)
(\zeta_{\sigma}-\delta N_\mathrm{ini})-4N_{\not{m}}}
+ (1-\bar{f}_{\sigma}) e^{4(\zeta_r-\delta N_\mathrm{ini}-N_{\not{m}})}
&=&\frac{\Gamma^2}{\Gamma_0^2}.\nonumber\\
\end{eqnarray}
For our calculation, we will neglect $\zeta_\sigma$, $\zeta_r$ and
$\delta N_\mathrm{ini}$.

\subsection{First order}
From Eq.(\ref{motonoshiki2}), we find at first order
\begin{eqnarray}
2\frac{\delta
 \Gamma^{(1)}_m}{\Gamma_0}&=&-(4+\epsilon_n)
\bar{f}_\sigma N_m^{(1)}-4(1-\bar{f}_\sigma)N_m^{(1)}\\
2\frac{\delta\Gamma^{(1)}_{\not{m}}}{\Gamma_0}
&=&-4 \bar{f}_\sigma N_{\not{m}}^{(1)}
-4(1-\bar{f}_\sigma)N_{\not{m}}^{(1)}.
\end{eqnarray}
Solving the above equations, we find
\begin{eqnarray}
N^{(1)}_m&=& -p_{\sigma,n} \frac{\delta \Gamma^{(1)}_m}{\Gamma_0}\\
N_{\not{m}}^{(1)}&=& -\frac{1}{2} \frac{\delta
 \Gamma^{(1)}_{\not{m}}}{\Gamma_0}
\end{eqnarray}
where the coefficient is defined by
\begin{eqnarray}
p_{\sigma,n}&\equiv& \frac{2(\bar{\rho}_{\sigma,0}+\bar{\rho}_{r,0})}
{(4+\epsilon_n)\bar{\rho}_{\sigma,0}+4\bar{\rho}_{r,0}}.
%\nonumber\\
%r_\sigma &\equiv& \frac{(4+\epsilon_n) \bar{\rho}_{\sigma,0}}
%{(4+\epsilon_n) \bar{\rho}_{\sigma,0}+4 \bar{\rho}_{r,0}}.
\end{eqnarray}
Therefore, the curvature perturbation created by the modulation is
given by
\begin{eqnarray}
\delta N^{(1)}&\equiv& N_m^{(1)}-N_{\not{m}}^{(1)}\nonumber\\
&=& \left(-p_{\sigma,n}+\frac{1}{2}\right)
\frac{\delta\Gamma^{(1)}}{\Gamma_0}.
\end{eqnarray}
Obviously, generation of the curvature perturbation is possible when
$\epsilon_n\ne 0$ (i.e. when two components ($\rho_\sigma$ and
$\rho_r$) are distinguishable in
their scaling relations).

\subsection{$f_{NL}$}
Again, we find for the second order perturbations
\begin{eqnarray}
N_m^{(2)}&=& p_{\sigma,n}
\left(\frac{(4+\epsilon_n)^2\bar{f_\sigma}}{2}+8(1-\bar{f}_\sigma)\right)
\left(N_m^{(1)}\right)^2 \nonumber\\
&&
-p_{\sigma,n}
 \frac{\left(\delta \Gamma^{(1)}\right)^2+\Gamma_0\delta \Gamma^{(2)}}
{\Gamma_0^2}\nonumber\\
N_{\not{m}}^{(2)}&=& \frac{1}{2}
\left[8\left(N_{\not{m}}^{(1)}\right)^2
-\frac{\left(\delta \Gamma^{(1)}\right)^2+\Gamma_0\delta \Gamma^{(2)}}
{\Gamma_0^2}\right].
\end{eqnarray}
Using the relations
\begin{eqnarray}
N_m^{(1)}&=&-p_{\sigma,n} \frac{\delta \Gamma^{(1)}}{\Gamma_0}\nonumber\\
&=&- \frac{2p_{\sigma,n}}{1-2p_{\sigma,n}}\delta N^{(1)}\\
N_{\not{m}}^{(1)}&=&-\frac{1}{2}\frac{\delta \Gamma^{(1)}}{\Gamma_0}\nonumber\\
&=&- \frac{1}{1-2p_{\sigma,n}}\delta N^{(1)},
\end{eqnarray}
and the definition (\ref{gammma2def}), we find 
\begin{eqnarray}
N_m^{(2)}&=& \left[
\left\{2(4+\epsilon_n)^2\bar{f}_\sigma+32(1-\bar{f}_\sigma)\right\}
\frac{p_{\sigma,n}^3}{(1-2p_{\sigma,n})^2}\right.\nonumber\\
&&\left.-\frac{4p_{\sigma,n}(1+A)}{(1-2p_{\sigma,n})^2}\right]\left(\delta
N^{(1)}\right)^2
\nonumber\\
N_{\not{m}}^{(2)}&=& \left[
\frac{4}{(1-2p_{\sigma,n})^2}
-\frac{2(1+A)}{(1-2p_{\sigma,n})^2}\right]\left(\delta
N^{(1)}\right)^2.\nonumber\\
\end{eqnarray}
We thus find 
\begin{eqnarray}
f_{NL}&=&\frac{5}{3}
\left[\frac{p_\sigma^3 (4+\epsilon_n)^2\bar{f}_\sigma
+16p_\sigma^3(1-\bar{f}_\sigma)-2p_\sigma-1}
{(1-2p_\sigma)^2}
\right]\nonumber\\
&&+\frac{5A}{3}\frac{1}{1-2p_\sigma},
\end{eqnarray}
where the last term gives the A-dependent contribution.
The single-component limit is given by
$\bar{f}_\sigma \rightarrow 1$ and
$p_\sigma\rightarrow 2/(4+\epsilon_n)$, where one may find significant
non-Gaussianity;
\begin{eqnarray}
f_{NL}&=&-\frac{5(4+\epsilon_n)}{3\epsilon_n}
-5A,
\end{eqnarray}
which shows that the sign of the first term (A-independent
contribution) is determined by $\epsilon_n$.
We find positive sign for $\epsilon_n<0$,
while it goes negative when $\epsilon_n>0$.
Interestingly,  neither $\Gamma(\varphi)$ nor
$g(\varphi)$ are responsible for the first term $f_{NL}\propto -
1/\epsilon_n$, which may become large even though $\rho_\sigma$
is dominating the Universe.

In the opposite limit, $\bar{f}_\sigma\rightarrow 0$ and
$p_\sigma\rightarrow 1/2$, we find 
\begin{equation}
f_{NL}\propto
\frac{1}{(1-2p_\sigma)^2} \gg 1,
\end{equation}
as expected.

\section{Partial decay}
More practically, there could be a moment when a fraction of the matter
component decays modulated and the decaying component does not have
significant interaction with the remaining (matter) components.
This could be realized when the non-relativistic matter contains particles
that belong to the hidden sector.

For the multi-component Universe that contains both matter ($\rho_\sigma$
and $\rho_\Delta$) and radiation ($\rho_r$), the uniform density hypersurfaces defined
for the partial decay is given by
\begin{eqnarray}
\rho_{\sigma,\Gamma}+
\rho_{\Delta,\Gamma} +\rho_{r,\Gamma} &\equiv& 3M_p^2 \Gamma^2,
\end{eqnarray}
where $\rho_{\sigma,\Gamma}$, $\rho_{\Delta,\Gamma}$ and
$\rho_{r,\Gamma}$ are the energy densities of the components at
$H=\Gamma$ ($\Gamma$ is the decay rate of the component $\rho_\Delta$).

Ignoring component perturbations ($\zeta_i\simeq 0$) and 
the initial perturbation ($\delta N_\mathrm{ini}\simeq 0$), we find
\begin{eqnarray}
\bar{f}_\sigma e^{-3N_m}
+\bar{f}_\Delta e^{-3N_m}
 + (1-f_\sigma-\bar{f}_\Delta) e^{-4N_m}
&=&\frac{\Gamma^2}{\Gamma_0^2}\nonumber\\
\bar{f}_\sigma e^{-3N_{\not{m}}}
+\bar{f}_\Delta e^{-4N_{\not{m}}}
 + (1-f_\sigma-\bar{f}_\Delta) e^{-4N_m}
&=&\frac{\Gamma^2}{\Gamma_0^2},\nonumber\\
\end{eqnarray}
where the coefficients are defined by
\begin{eqnarray}
\bar{f}_\sigma &\equiv& \frac{\bar{\rho}_{\sigma,0}}
{\bar{\rho}_{\sigma,0} +\bar{\rho}_{\Delta,0} 
+\bar{\rho}_{r,0}}\\
\bar{f}_\Delta &\equiv& \frac{\bar{\rho}_{\Delta,0}}
{\bar{\rho}_{\sigma,0} +\bar{\rho}_{\Delta,0} 
+\bar{\rho}_{r,0}}.
\end{eqnarray}
As before, the subscript ``0'' is used to define the quantities at $H=\Gamma_0$.

\subsection{First order}
We find at first order
\begin{eqnarray}
2\frac{\delta \Gamma^{(1)}}{\Gamma_0}&=&-3\bar{f}_\sigma N_m^{(1)}
-3\bar{f}_\Delta N_m^{(1)}\nonumber\\
&&-4(1-\bar{f}_\sigma-\bar{f}_\Delta)N_m^{(1)}\\
2\frac{\delta\Gamma^{(1)}}{\Gamma_0}
&=&-3\bar{f}_\sigma N_m^{(1)}
-4\bar{f}_\Delta N_m^{(1)}\nonumber\\
&&-4(1-\bar{f}_\sigma-\bar{f}_\Delta)N_m^{(1)}
\end{eqnarray}
Solving the above equations, we find
\begin{eqnarray}
N^{(1)}_m&=& -p_{\Delta} \frac{\delta \Gamma^{(1)}}{\Gamma_0}\\
N_{\not{m}}^{(1)}&=& -\not{p}_{\Delta} \frac{\delta
 \Gamma^{(1)}}{\Gamma_0},
\end{eqnarray}
where the coefficients are defined by
\begin{eqnarray}
p_{\Delta}&\equiv& \frac{2(\bar{\rho}_{\sigma,0}+\bar{\rho}_{\Delta,0}
+\bar{\rho}_{r,0})}
{3\bar{\rho}_{\sigma,0}+3\bar{\rho}_{\Delta,0}+
4\bar{\rho}_{r,0}}\\
\not{p}_{\Delta}&\equiv& \frac{2(\bar{\rho}_{\sigma,0}+\bar{\rho}_{\Delta,0} 
+\bar{\rho}_{r,0})}
{3\bar{\rho}_{\sigma,0}+4\bar{\rho}_{\Delta,0}+
4\bar{\rho}_{r,0}}.
\end{eqnarray}
Therefore, the curvature perturbation created by the modulation is
\begin{eqnarray}
\delta N^{(1)}&\equiv& N_m^{(1)}-N_{\not{m}}^{(1)}\nonumber\\
&=& \left(-p_{\Delta}+\not{p}_{\Delta}\right)
\frac{\delta\Gamma^{(1)}}{\Gamma_0}\nonumber\\
&\simeq&
-p_\Delta r_\Delta
\frac{\delta\Gamma^{(1)}}{\Gamma_0},
\end{eqnarray}
where the last approximation is valid when $r_\Delta \ll 1$.
Here the coefficient is defined by
\begin{equation}
r_\Delta\equiv\frac{\bar{\rho}_{\Delta,0}}
{3\bar{\rho}_{\sigma,0}+3\bar{\rho}_{\Delta,0}+
4\bar{\rho}_{r,0}}.
\end{equation}

\subsection{Second order}
We find for the second order perturbations
\begin{eqnarray}
N_m^{(2)}&=& 
p_\Delta\left(8-\frac{7}{2}(\bar{f}_\sigma+\bar{f}_\Delta)\right)\left(N_m^{(1)}\right)^2\nonumber\\
&&-p_\Delta\frac{\left(\delta \Gamma^{(1)}\right)^2+\Gamma_0\delta \Gamma^{(2)}}{\Gamma_0^2}\\
N_{\not{m}}^{(2)}&=& 
{\not{p}}_\Delta\left(8-\frac{7}{2}\bar{f}_\sigma\right)\left(N_m^{(1)}\right)^2\nonumber\\
&&-{\not{p}}_\Delta\frac{\left(\delta \Gamma^{(1)}\right)^2+\Gamma_0\delta \Gamma^{(2)}}{\Gamma_0^2}.
\end{eqnarray}
Using the relations
\begin{eqnarray}
N_m^{(1)}&=&-p_\Delta \frac{\delta \Gamma^{(1)}}{\Gamma_0}\nonumber\\
&=&\frac{p_\Delta}{p_\Delta-{\not{p}}_\Delta}\delta N^{(1)}\\
N_{\not{m}}^{(1)}&=&
-{\not{p}}_\Delta \frac{\delta \Gamma^{(1)}}{\Gamma_0}\nonumber\\
&=&\frac{{\not{p}}_\Delta}{p_\Delta-{\not{p}}_\Delta}\delta N^{(1)},
\end{eqnarray}
and the definition (\ref{gammma2def}), we find 
\begin{eqnarray}
N_m^{(2)}&=&
 \left[\frac{p_\Delta^3(16-7\bar{f}_\sigma-7\bar{f}_\Delta)}
{2(p_\Delta-{\not{p}}_\Delta)^2}
-\frac{p_\Delta(1+A)}{(p_\Delta-{\not{p}}_\Delta)^2}\right]\left(\delta
N^{(1)}\right)^2
\nonumber\\
N_{\not{m}}^{(2)}&=& \left[
\frac{{\not{p}}_\Delta^3(16-7\bar{f}_\sigma)}
{2(p_\Delta-{\not{p}}_\Delta)^2}
-\frac{{\not{p}}_\Delta(1+A)}{(p_\Delta-{\not{p}}_\Delta)^2}\right]\left(\delta
N^{(1)}\right)^2.
\end{eqnarray}
Then, $f_{NL}$ is calculated from 
\begin{eqnarray}
f_{NL}&=&\frac{5}{6}\frac{N_m^{(2)}-N_{\not{m}}^{(2)}}{\left(\delta
N^{(1)}\right)^2}.
\end{eqnarray}

\section{Conclusion and discussion}

The early Universe after inflation may have many components labeled by
the density $\rho_i$ and each component may have distinguishable scaling
relation $\rho_i \propto a^{k_i}$.  
They could be oscillations, topological defects, relativistic and
non-relativistic particles.
If those components are decaying into radiation in the end, there could be
a generation of the curvature perturbation.
In this paper, the mechanism of the modulated decay has been considered
for the multi-component Universe. 
The conventional ``modulated reheating'' scenario is realized in the
single-component Universe.

In this paper we found the basic formulation,
which is useful in calculating modulated decays in the
 multi-component Universe.
We have found useful results, in which the non-Gaussianity parameter is
separated into A-dependent and A-independent terms.
Here $A$ is determined by the form of $\Gamma(\varphi)$ and the
evolution function $g(\varphi_*)$.
Interestingly, $f_{NL}$ may appear with either positive or negative signs.
 We found that the component, whose scaling is similar to
the radiation ($k_i \sim -4$), will generate
 significant non-Gaussianity in the single-component 
(conventional reheating) limit.
In that way, the {\bf conventional modulated reheating}
 caused by the oscillation
 may {\it crucially
depend on the amplitude} at the decay. 
For instance, consider the potential
for the oscillations given by
\begin{equation}
V(\sigma)\simeq \frac{1}{2}m^2\sigma^2+\frac{\lambda_4}{4}\sigma^4
+\frac{\lambda_6}{6}\frac{\varphi^6}{M_p^2}.
\end{equation}
If the oscillations decay when $\varphi^6$ is dominant,
one will find $f_{NL}<0$. 
If the oscillations decay when $\varphi^4$ is dominant,
one will find $|f_{NL}|\gg 1$, where the sign could be either positive
or negative. 
The scaling of the density changes during the oscillations.
One will find conventional result when the quadratic term is
dominating.
In the intermediate region one may find the density scaling $\rho_\sigma \propto
a^{-k_\sigma}$, where (effectively) $3\le k_\sigma\le 6$ is possible.
As the result, in the practical calculation the curvature perturbation
and the non-Gaussianity may
depend crucially on the amplitude of the oscillations, even if the decay
occurs in the single-component Universe.

{\noindent {\bf Note added}}: While finalizing this paper, we found a
couple of papers~\cite{noteadded} which has some
overlaps with our models.
In the appendix we are discussing the correspondences between these works.

\section{Acknowledgment}
T.M wishes to thank K.~Shima for encouragement, and his colleagues at
Nagoya university and Lancaster university for their kind hospitality
and many invaluable discussions.
This work is supported in part by Grant-in-Aid for Scientific research
from the Ministry of Education, Sci- ence, Sports, and Culture (MEXT),
Japan, No. 21111006, No. 22244030, and No. 23540327 (K.K.).
S.E. is supported by the Grant-in-Aid for Nagoya University Global COE Program,
"Quest for Fundamental Principles in the Universe: from Particles to the Solar
System and the Cosmos".

\appendix
\section{Non-Linear formalism and the curvaton mechanism}
In this appendix, we first review the basics of the curvaton mechanism
in the light of the non-linear formalism,
and then compare our results with Ref.\cite{noteadded}.
Notations in Ref.\cite{noteadded} are discriminated using the subscripts
``LT-A'', when necessary.

The non-linear formalism in the curvaton mechanism is given by the formula
\begin{eqnarray}
\zeta_\sigma&=&\delta N + \frac{1}{3}\ln
 \left(\frac{\rho_\sigma}{\bar{\rho}_\sigma}\right),\\
\zeta_r&=&\delta N + \frac{1}{4}\ln
 \left(\frac{\rho_r}{\bar{\rho}_r}\right).
\end{eqnarray}
Here $\delta N$ is the perturbation of $N$, which is measured
between two hypersurfaces, which are usually the flat 
and the uniform density hypersurfaces.
Besides $\delta N$, we have to define the other quantities
($\rho_\sigma$, $\rho_r$) and ($\bar{\rho}_\sigma$, $\bar{\rho}_r$).
Those quantities are defined on the uniform density hypersurface for which
$\delta N$ is defined. 

We thus find for the uniform density hypersurface $H=H_A$;
\begin{eqnarray}
\zeta_{\sigma,A}&=&\delta N_A(t_A) + \frac{1}{3}\ln
 \left(\frac{\rho_{\sigma,A}(x,t_A)}{\bar{\rho}_{\sigma,A}(t_A)}\right),\\
\zeta_{r,A}&=&\delta N_A(t_A) + \frac{1}{4}\ln
 \left(\frac{\rho_{r,A}(x,t_A)}{\bar{\rho}_{r,A}(t_A)}\right).
\end{eqnarray}
Solving these equations we find
\begin{eqnarray}
\label{eq-o}
\rho_{\sigma,A} &=&\bar{\rho}_{\sigma,A}e^{3(\zeta_{\sigma,A}-\delta N_A)},\nonumber\\
\rho_{r,A} &=&\bar{\rho}_{r,A}e^{4(\zeta_{r,A}-\delta N_A)}.
\end{eqnarray}
The trivial identity is
\begin{equation}
\frac{\rho_{\sigma,A}+\rho_{r,A}}{\bar{\rho}_{\sigma,A}+\bar{\rho}_{r,A}}=1,
\end{equation}
where $\rho_{\sigma,A}$ and $\rho_{r,A}$ can be replaced using
Eq.(\ref{eq-o}).
We find the equation
\begin{eqnarray}
\bar{f}_{\sigma,A} e^{3(\zeta_{\sigma,A}-\delta N_A)}+
(1-\bar{f}_{\sigma,A}) e^{4(\zeta_{r,A}-\delta N_A)}&=&1,
\end{eqnarray}
where the ratio is defined by
\begin{eqnarray}
\bar{f}_{\sigma,A}&=&
 \frac{\bar{\rho}_{\sigma,A}}{\bar{\rho}_{\sigma,A}+\bar{\rho}_{r,A}}.
\end{eqnarray}
We find at first order
\begin{eqnarray}
\delta N_A&=& r_{\sigma,A}\zeta_{\sigma,A}
 +(1-r_{\sigma,o})\zeta_{r,o}\nonumber\\
&=& \delta N_A \nonumber\\
&&+ \frac{r_A}{3}\ln
 \left(\frac{\rho_{\sigma,A}}{\bar{\rho}_{\sigma,A}}\right)
+\frac{(1-r_A)}{4}\ln
 \left(\frac{\rho_{r,A}}{\bar{\rho}_{r,A}}\right).\nonumber\\
\end{eqnarray}
The trivial identity is
\begin{eqnarray}
\frac{r_A}{3}\ln
 \left(\frac{\rho_{\sigma,A}}{\bar{\rho}_{\sigma,A}}\right)
+\frac{(1-r_A)}{4}\ln
 \left(\frac{\rho_{r,A}}{\bar{\rho}_{r,A}}\right)&=&0.
\end{eqnarray}
For the expansion $\delta \rho_i\equiv \rho_i-\bar{\rho}_i$,
 the above equation gives the obvious identity
\begin{eqnarray}
\delta \rho_{\sigma,A}+\delta \rho_{r,A}&=&0.
\end{eqnarray}

\begin{figure}[t]
\centering
\includegraphics[width=1.0\columnwidth]{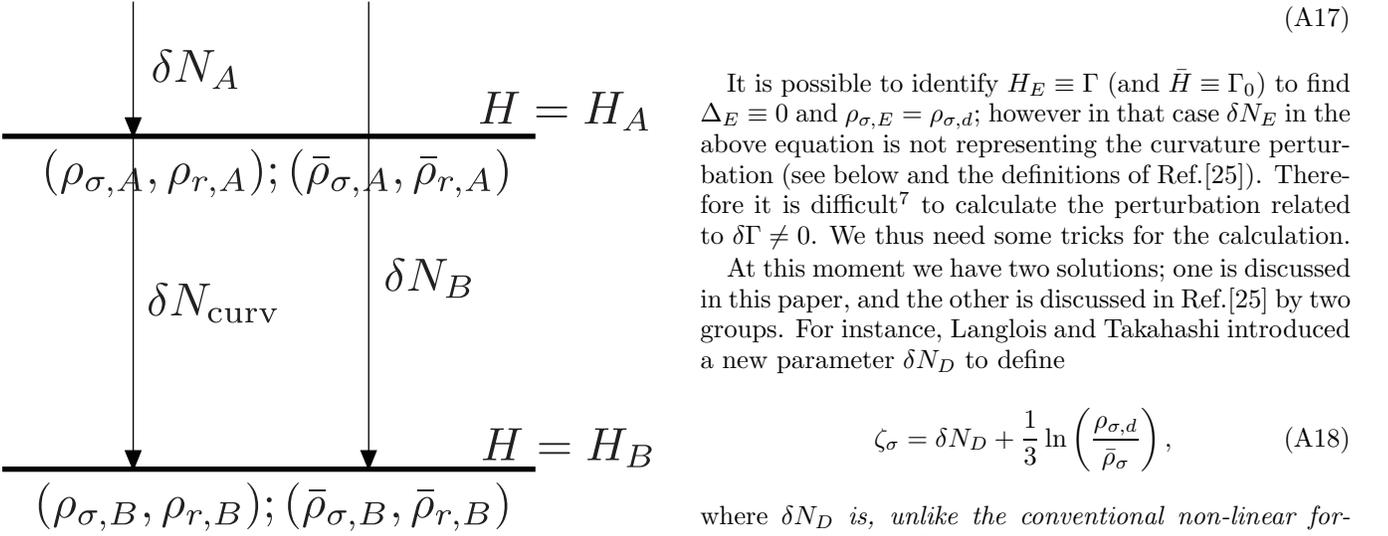}
 \caption{Definitions of the component perturbations are illustrated.
For instance, $\zeta_{\sigma,A}\equiv \delta N_A + \frac{1}{3}\ln
 \left(\frac{\rho_{\sigma,A}}{\bar{\rho}_{\sigma,A}}\right)$ is defined using
$\delta N_A$ (measured from the flat hypersurfaces to $H=H_A$), 
$\rho_{\sigma,A}$ and
 its mean value $\bar{\rho}_{\sigma,A}$ (both are defined on $H=H_A$).} 
\label{fig:appendix1}
\end{figure}
One may evaluate the non-linear formalism {\bf away} from
$H=H_A$. (See Fig.\ref{fig:appendix1}.)
Choosing {\bf another} hypersurface $H=H_B$, one can evaluate
a similar equation
\begin{eqnarray}
\delta N_B&=& r_{\sigma,B}\zeta_{\sigma,B}
 +(1-r_{\sigma,B})\zeta_{r,B}\nonumber\\
&=& r_{\sigma,B}\zeta_{\sigma,A}
 +(1-r_{\sigma,B})\zeta_{r,A},
\end{eqnarray}
where the constancy of the component perturbations
($\zeta_{i,A}=\zeta_{i,B}$) has been used.

Note that $\delta N_\mathrm{curv}\equiv 
\delta N_B-\delta N_A$ gives the ``evolution of $\delta N$'' between the two
hypersurfaces $H_A$ and $H_B$.
We thus find for $r_{\sigma,B}\gg r_{\sigma,A}$:
\begin{eqnarray}
\label{curv-appex}
\delta N_\mathrm{curv}
&=& (r_{\sigma,B}-r_{\sigma,A})\zeta_{\sigma,A}
-(r_{\sigma,B}-r_{\sigma,A})\zeta_{r,A}\nonumber\\
&\simeq&
r_{\sigma,B}\left[\frac{\delta \rho_{\sigma,A}}{3\bar{\rho}_{\sigma,A}}\right].
\end{eqnarray}
Note that $\delta N_A$ does not appear in $\delta N_\mathrm{curv}$
because of the obvious cancellation (see Fig.\ref{fig:appendix1}). 

If one defines $H_A$ at the beginning of the curvaton
oscillation and $H_B$ at the decay, $\delta N_\mathrm{curv}$ gives the evolution
of the curvature perturbation in the conventional curvaton mechanism.

The conventional curvature perturbation generated by the primordial
inflation can be included as $\delta N_\mathrm{inf}\simeq \delta N_A$.

\subsection{Why difficult?}

The formalism that can be applied for the modulation at the end of the
curvaton mechanism has
already been discussed by Enomoto-Kohri-Matsuda(EKM) in Ref.\cite{EKM}.
When the curvaton decay ($\rho_\sigma$ decay)
is modulated, the non-linear formalism just
after the decay ($t=t_E$) can be separated as
\begin{eqnarray}
\zeta_{\sigma,\mathrm{E}}&=&\delta N_\mathrm{E}+
\frac{1}{3}\ln \left(\frac{\rho_{\sigma,d}}
{\bar{\rho}_{\sigma}}\right)+
\frac{1}{4}\ln \left(\frac{\rho_{\sigma,\mathrm{E}}}
{\rho_{\sigma,d}}\right)\nonumber\\
&=&\delta N_\mathrm{E}+
\frac{1}{3}\ln \left(\frac{\rho_{\sigma,\mathrm{E}}}
{\bar{\rho}_{\sigma}}\right)+\Delta_E
\nonumber\\
\zeta_{r,\mathrm{E}}&=&\delta N_\mathrm{E} + \frac{1}{4}\ln
 \left(\frac{\rho_{r,\mathrm{E}}}{\bar{\rho}_{r}}\right),
\end{eqnarray}
where we defined
\begin{equation}
\Delta_\mathrm{E}\equiv\frac{1}{12}\ln \left(\frac{\rho_{\sigma,d}}
{\rho_{\sigma,\mathrm{E}}}\right).
\end{equation}
Here the subscript ``E'' means that the quantities
are evaluated at $t=t_\mathrm{E}$;
$\rho_{\sigma,E}(x,t_E)$ is the inhomogeneous density of the curvaton
remnant (radiation density separated from the total density of the radiation) and $\rho_{\sigma,d}(x)$ is the density when
$\rho_\sigma$ decays. 
We have chosen the ordering $\bar{\rho}_{\sigma}\ge\rho_{\sigma,d}
\ge \rho_{\sigma,E}$ just for simplicity.

Again, the trivial identity
\begin{equation}
\frac{\rho_{\sigma,\mathrm{E}}+\rho_{r,\mathrm{E}}}
{\bar{\rho}_{\sigma}+\bar{\rho}_{r}}=\frac{H^2_\mathrm{E}}{\bar{H}^2}
\end{equation}
gives
\begin{eqnarray}
\bar{f}_{\sigma} e^{3(\zeta_{\sigma,\mathrm{E}}-\delta N_\mathrm{E}
-\Delta_\mathrm{E})}
+(1-\bar{f}_{\sigma}) e^{4(\zeta_{r}-\delta N_\mathrm{E})}
&=&\frac{H^2_\mathrm{E}}{\bar{H}^2}.\nonumber\\
\end{eqnarray}

It is possible to identify $H_E\equiv\Gamma$ (and $\bar{H}\equiv
\Gamma_0$) to find $\Delta_E\equiv 0$ and
$\rho_{\sigma,E}=\rho_{\sigma,d}$; however in that case
$\delta N_E$ in the above equation is not representing the curvature
perturbation 
 (see below and the definitions of Ref.\cite{noteadded}).
Therefore it is difficult\footnote{This is our personal impression.
A reader might be able to find more convincing way of calculation
without using redefinitions of the quantities. Another way
of calculation can be found in Ref.\cite{Chiamin-mod}, in which the
definitions of the quantities could be more straight than the previous
papers.}
 to calculate the perturbation related to $\delta \Gamma\ne 0$.
We thus need some tricks for the calculation.

At this moment we have two solutions; one is discussed in this paper,
and the other is discussed in 
Ref.\cite{noteadded} by two groups.
For instance, Langlois and Takahashi introduced a new parameter
$\delta N_D$ to define 
\begin{equation}
\zeta_{\sigma}=\delta N_D + \frac{1}{3}\ln
 \left(\frac{\rho_{\sigma,d}}{\bar{\rho}_{\sigma}}\right),
\end{equation}
where {\it $\delta N_D$ is, unlike the conventional non-linear
formalism, not identified with 
the curvature perturbation $\zeta$, while $\zeta_\sigma$
is identical to the conventional component perturbation.}
These definitions are obviously strange when they are compared with the
normal definitions.
Also, it could be rather difficult to understand why the above
definition of $\zeta_\sigma$ is identical to the normal definition.
{\bf In our paper, 
we have introduced fundamental quantities $N_m$ and $N_{\not{m}}$
defined in the separate Universe, which
can be used to calculate $\delta N\equiv N_{\not{m}}-N_m$.
Note that our definitions are simply explaining $\delta N$ in the
separate Universe hypothesis.}
Below, we will take a closer look at these definitions.

\subsection{Quantities defined in Ref.\cite{noteadded}}

We are going to show obvious correspondences between
quantities defined in Ref.\cite{noteadded} and ours
in Fig.\ref{fig:mod-uni}. 
Let us consider the ``simplest multi-component Universe'' that has been
defined in this paper, which is the Universe whose density consists of matter
$\rho_\sigma$ and radiation $\rho_r$.
This model is familiar among the conventional curvaton models.
The decay of the matter is therefore looks like a curvaton decay.
It is possible to calculate the mixed (modulation-curvaton)
perturbations when the curvaton perturbations are not negligible, however
in the main part of this paper we have been focusing on the scenario in
which modulation is 
dominating the cosmological perturbation.

\begin{figure}[t]
\centering
\includegraphics[width=1.0\columnwidth]{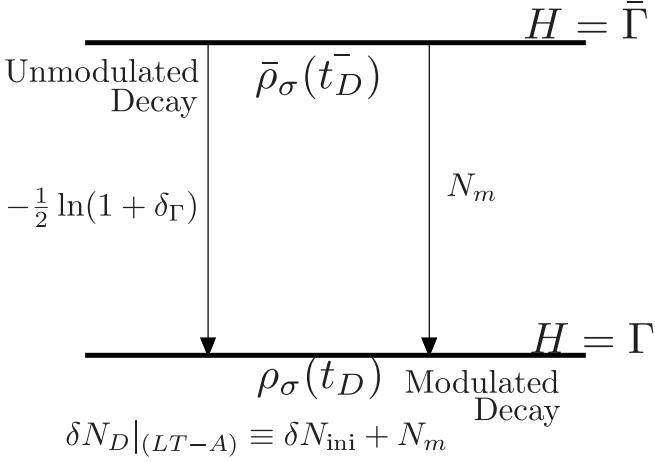}
 \caption{Unlike the conventional definition of the non-linear
 formalism, component perturbations of ref.\cite{noteadded} are defined 
using $\rho_\sigma$ on $H=\Gamma$ and 
 $\bar{\rho}_\sigma$ on $H=\bar{\Gamma}$.
As a result, $-\frac{1}{2}\ln(1+\delta_\Gamma)$ has to be 
 subtracted from  $\delta N_D$ to get the conventional $\delta N$.}
\label{fig:appendix2}
\end{figure}

In Ref.\cite{noteadded}, they have defined the non-linear formalism
\begin{equation}
\zeta_\sigma=\delta N_D + \frac{1}{3}\ln
\left(\frac{\rho_\sigma(t_D)}{\bar{\rho}_\sigma}\right),
\end{equation}
where the hypersurface defined by $t_D$ ($H=\Gamma$) 
is modulated.
In that case $\delta N_D$ cannot represent the usual curvature perturbation.
(See Fig.\ref{fig:appendix2}.)

In our formalism, these quantities are corresponding to
\begin{eqnarray}
\delta N_D|_{(LT-A)}&\leftrightarrow& \delta N_\mathrm{ini}+N_m\nonumber\\
\rho_\sigma(t_D)|_{(LT-A)}&\leftrightarrow&\rho_{\sigma, \Gamma,m}
=\rho_{\sigma,0}e^{-3N_m}\nonumber\\
\bar{\rho}_\sigma(\bar{t}_D)|_{(LT-A)}&\leftrightarrow&\bar{\rho}_{\sigma,0}.
\end{eqnarray}
We thus find
\begin{equation}
\zeta_\sigma|_{(LT-A)}\leftrightarrow\zeta_\sigma.
\end{equation}
They also defined
\begin{equation}
(1+\delta_\Gamma)^2\equiv \Gamma^2/\bar{\Gamma}^2,
\end{equation}
which gives the correspondence
 \begin{equation}
(1+\delta_\Gamma)^2\leftrightarrow e^{-4N_{\not{m}}}.
\end{equation}
Finally, they have defined the post-decay curvature perturbation
\begin{equation}
\zeta=\delta N_D +\frac{1}{2}\ln(1+\delta_\Gamma),
\end{equation}
which corresponds to
\begin{equation}
\zeta|_{(LT-A)}\leftrightarrow \delta N_{\mathrm{ini}} +N_m
 -N_{\not{m}}.
\end{equation}
In our calculation the curvature perturbation generated by the modulated
decay is given by
\begin{equation}
\delta N_{\mathrm{mod}} \equiv N_m-N_{\not{m}}.
\end{equation}
Therefore, the correspondence is obvious between our calculation and
Ref.\cite{noteadded}. 

In finding the curvaton contribution they evaluated
\begin{equation}
\zeta=\zeta_r -\frac{r}{6}\delta_\Gamma + \frac{r}{3}S,
\end{equation}
where the first and the last terms are originally given by
\begin{equation}
\label{above-eq}
\zeta_r + \frac{r}{3}S=
r\zeta_\sigma + (1-r)\zeta_r\equiv \delta N_\mathrm{ini},
\end{equation}
where $\delta N_\mathrm{ini}$ is defined previously in this paper.
If the curvaton hypothesis is valid and the component perturbations are
constant, one may evaluate the component perturbation at
$H=H_\mathrm{osc}$ as
\begin{eqnarray}
\zeta_\sigma=\zeta_\sigma(t_\mathrm{osc})
=\delta N_\mathrm{inf}+\frac{1}{3}\ln\frac{\rho_\sigma(t_\mathrm{osc})}
{\bar{\rho}_\sigma(t_\mathrm{osc})},
\end{eqnarray}
where $\delta N_\mathrm{inf}$ denotes the curvature perturbation just
at the beginning of the oscillation.
Substituting the component perturbations (defined at $H_\mathrm{osc}$)
 into the above equation (\ref{above-eq}), one will find
\begin{eqnarray}
\zeta_r + \frac{r}{3}S&=&
r\zeta_\sigma(t_\mathrm{osc}) + (1-r)
\zeta_r(t_\mathrm{osc})\nonumber\\
&=&\delta N_\mathrm{inf} +
 \frac{r}{3}\ln\frac{\rho_\sigma(t_\mathrm{osc})} 
{\bar{\rho}_\sigma(t_\mathrm{osc})}\nonumber\\
&&+ \frac{1-r}{4}\ln\frac{\rho_r(t_\mathrm{osc})} 
{\bar{\rho}_r(t_\mathrm{osc})}.
\end{eqnarray}
Although a deformation is needed, it is easy to find that the result
is consistent with Eq.(\ref{curv-appex}).

Using the above formula, they started perturbation with regard to
the perturbation of $S$.
For instance, Langlois and Takahashi considered
for the ``curvaton perturbation''
\begin{equation}
S\equiv 3(\zeta_\sigma-\zeta_r)=\frac{2\sigma}{\sigma}-\frac{\delta \sigma^2}{\sigma^2}
+\frac{2}{3}\frac{\delta \sigma^3}{\sigma^3},
\end{equation}
and for the ``inflaton perturbation''
\begin{equation}
\zeta_r=\frac{H}{\dot{\phi}}\delta \phi \simeq \delta N_\mathrm{inf}.
\end{equation}
These definitions are based on the usual
curvaton hypothesis (i.e, valid when one can disregard $\delta
\rho_r/\rho_r$).

\end{document}